\documentstyle[epsfig,aps,floats,preprint]{revtex}

\newcommand{\beq}{\begin{equation}}
\newcommand{\eeq}{\end{equation}}
\newcommand{\bea}{\begin{eqnarray}}
\newcommand{\eea}{\end{eqnarray}}

\def\laq{~\raise 0.4ex\hbox{$<$}\kern -0.8em\lower 0.62
ex\hbox{$\sim$}~}
\def\gaq{~\raise 0.4ex\hbox{$>$}\kern -0.7em\lower 0.62
ex\hbox{$\sim$}~}

\def \ra {\rightarrow}

\def \Da {\Delta}

\def \a {\alpha}

\def \Ga {\Gamma}
\def \ga {\gamma}
\def \sg {\sigma}
\def \da {\delta}

\def \r {\rho}

\def \fb {\overline \phi}
\def \rb {\overline \rho}
\def \sgb {\overline \sg}
\def \pb {\overline p}
\def \fbp {\dot{\fb}}

\begin{document}
\par
\begingroup
%\twocolumn[%

\begin{flushright}
BA-TH/01-416\\
hep-th/0109137\\
\end{flushright}
\vskip 1true cm

\vspace{15mm}
{\large\bf\centering\ignorespaces
Low-energy graceful exit\\ 
in anisotropic string cosmology backgrounds
\vskip2.5pt}

\bigskip
{\dimen0=-\prevdepth \advance\dimen0 by23pt
\nointerlineskip \rm\centering
\vrule height\dimen0 width0pt\relax\ignorespaces
G. De Risi${}^{(1)}$ and M. Gasperini${}^{(2,3)}$
\par}
\bigskip
{\small\it\centering\ignorespaces
${}^{(1)}$
Dipartimento di Fisica, Universit\`a di Perugia, 
Via A. Pascoli, 06123 Perugia, Italy \\
${}^{(2)}$
Dipartimento di Fisica, Universit\`a di Bari, 
Via G. Amendola 173, 70126 Bari, Italy \\
${}^{(3)}$
Istituto Nazionale di Fisica Nucleare, Sezione di Bari,
Bari, Italy \\
\par}

\par
\bgroup
\leftskip=0.10753\textwidth \rightskip\leftskip
\dimen0=-\prevdepth \advance\dimen0 by17.5pt \nointerlineskip
\small\vrule width 0pt height\dimen0 \relax

\begin{abstract}
We discuss the possibility of a smooth transition from the pre- to the
post-big bang regime, in the context of the lowest-order string
effective action (without higher-derivative corrections), taking into
account with a phenomenological model of source the repulsive
gravitational effects due to the back-reaction of the quantum
fluctuations outside the horizon. We determine a set of necessary
conditions for a successful and realistic transition, and we find that
such conditions can be satisfied (by an appropriate model of source),
provided the background is higher-dimensional and anisotropic.

\end{abstract}

\vspace{5mm}
\begin{center}
---------------------------------------------\\
\vspace {5 mm}

To appear in {\bf Phys. Lett. B}
\end{center}

%\vspace{5mm}

\par\egroup
%\vskip2pc]
\thispagestyle{plain}
\endgroup


%\section {Introduction}
%\label{I}

According to the pre-big bang cosmological scenario \cite{1}, inspired
by the duality symmetries of the string effective action \cite{2}, and
also  recently motivated by models of brane-world dynamics \cite{3},
the present Universe is assumed to emerge from an inital state of very
low curvature and small couplings (in string units), asymptotically
approaching the string perturbative vacuum. The ``birth 
" of our Universe, in this context, may thus be represented as a process
of decay of the string perturbative vacuum, and described in the
language of quantum string cosmology as a transition between
the pre- and post-big bang regimes \cite{4,5} associated to a tunelling
(or anti-tunnelling \cite{6}) of the Wheeler-De Witt wave function in
minisuperspace. 

At a classical level the representation of this transition
process is problematic, as it requires a smooth evolution of the
background from an initial accelerated configuration in which the
curvature and the string coupling (i.e. the dilaton) are growing, to a
final decelerated configuration in which the curvature is decreasing,
and the dilaton is constant or decreasing -- the so-called ``graceful
exit". This requires, in particular, the regularization of the curvature
singularities which in general affect the cosmological solutions
of the string effective action and which disconnect, classically, the
duality-related pre- and post-big bang regimes. This also implies that
the growth of the dilaton has to be stopped, to avoid that the
curvature is regular in a frame but blows up in a different, conformally
related frame \cite{7}. 

For the lowest order gravi-dilaton string effective action 
there are  indeed ``no-go theorems" \cite{8}, excluding a smooth
transition even in the presence of a (local) dilaton potential and of
matter sources in the form of perfect fluids and/or Kalb-Ramond axions.
For such a reason, it has been repeatedly stressed, in the literature, the
need for including higher-order (quantum loops \cite{9,10} and
higher-derivative \cite{11} -\cite{13}) corrections in the string effective
action,  in order to smooth out the background singularities, and to
implement a graceful exit from the phase of pre-big bang inflation to the
subsequent phase of standard, decelerated evolution. 

The higher-derivative terms, in particular, can efficiently stop the
growth of the curvature during a phase of linear dilaton evolution
\cite{11}, thus preparing the background to the action of the loop
corrections, which in turn provide the necessary ``repulsive gravity"
effects \cite{10} needed to evade the classical singularity theorems
(see for instance \cite{14}), and to regularize the transition. 

The loop corrections, in fact, are physically induced by the
``back-reaction" of the quantum fluctuations against the classical
solution, which describes initially a  pre-big bang phase of growing
curvature and shrinking horizons. As the curvature is growing, the
quantum fluctuations are stretched outside the horizon, and it is
known that in this regime they are characterized by an effective
gravitational energy density which is negative \cite{15}, and which
may  favour the transition to the post-big bang branch of the classical
solution \cite{16}. Such a negative back-reaction is eventually damped
to zero when the curvature start decreasing, the horizon blows up
again, and all the fluctuations re-enter inside the horizon and in the
regime of positive energy density. It is important to notice, indeed, that
all successful examples of graceful exit (either with a non-local 
potential \cite{1,4}, higher-derivatives \cite{10,13}, or different
mechanisms  \cite{17}) always contain repulsive-gravity effects,
directly or  indirectly related to the quantum back-reaction of the loop
corrections. 

It should be recalled, at this point, that the mentioned no-go theorems,
formulated in the context of the lowest-order string effective action,
are all referred to a homogeneous and isotropic four-dimensional
background. If the isotropy and homogeneity assumptions are relaxed,
however, it is known that some singularities can be eliminated
(technically, ``boosted away") through an appropriate $O(d,d)$ duality
transformation, effective also at the tree-level \cite{18}. In that case,
the repulsive effects regularizing the singularities are due to the
antisymmetric tensor field introduced by the boost-transformation. Such
examples of regular backgrounds are not usually regarded as successful
models of graceful exit, however, because they describe a Universe that
after the transition is too inhomogeneous (see however \cite{19}), or
even contracting in all its dynamical dimensions \cite{20}, to be
realistic. 

The aim of this paper is to show, with an explicit example, that the
higher-derivative corrections are not at all necessary to formulate a
realistic model of graceful exit, which is homogeneous and which
contains, in its final configuration, three expanding dimensions. The
low-energy dynamics of the string effective action is enough, to this
purpose, provided the metric background is anisotropic, and provided
we take into account, with a phenomenological source term, the
repulsive gravitational effects due to the back-reaction of the
quantum fluctuations outside the horizon. 

We shall consider, in particular, a $D$-dimensional Bianchi I-type
metric background, with a time-dependent dilaton $\phi$, 
\beq
g_{\mu\nu}={\rm diag }(1, -a_i^2 \da _{ij}), ~~~~~
a_i=a_i(t), ~~~~~~ \phi=\phi(t), ~~~~~~ i=1, 2, \dots D-1,
\label{1}
\eeq
whose dynamical evolution is controlled by the low-energy
gravi-dilaton effective action: 
\beq 
S= -\int d^{D}x \sqrt{|g|}~
e^{-\phi}\left[R+ \left(\nabla \phi\right)^2\right] + \Ga (\phi, g, {\rm
matter})  
\label{2}
\eeq
(we are working in the string frame, and in units in which the string
tension $4 \pi \alpha'$ is set to unity). Here $\Ga$ is the effective action
for the matter fields, including the contribution of all the quantum
fluctuations, assumed to be subleading unless they are outside the
horizon. 

The variation of the action with respect to $g_{\mu\nu}$ and
$\phi$ leads to the equations of motion:
\bea
&&
R_{\mu\nu} -{1\over 2} g_{\mu\nu} R+ \nabla_\mu\nabla_\nu \phi
+{1\over 2}g_{\mu\nu} \left[\left(\nabla \phi\right)^2 -2 \nabla^2 \phi
\right]  ={1\over 2}e^\phi T_{\mu\nu}, \nonumber\\
&&
\left(\nabla \phi\right)^2 -2 \nabla^2 \phi -R=
e^\phi \sg ,
\label{3}
\eea
containing two source terms,
\beq
T_{\mu\nu}= {2\over \sqrt{-g}}{\da \Ga \over \da g^{\mu\nu}},
~~~~~~~~~~~ \sg={1\over \sqrt{-g}}{\da \Ga \over \da \phi}
\label{4}
\eeq
(i.e., the gravitational and dilatonic ``charge densities"). They are
assumed to be compatible with the isometries of the background
(\ref{1}), so that we can set
\beq
T_\mu\,^\nu=  {\rm diag} (\rho, -p_i^2 \da_i^j), ~~~~~~~\r=\r(t), 
~~~~~~~p_i=p_i(t), ~~~~~~~~ \sg=\sg(t). 
\label{5}
\eeq
We have thus $D+1$ independent equations, that can be cast in the
form (see for instance \cite{1,2}):
\bea
&&
\fbp^2 -\sum_iH_i^2= \rb e^{\fb} , \nonumber\\
&&
\dot H_i -H_i \fbp={1\over 2}(\pb_i + \sgb) e^{\fb}, \nonumber\\
&&
\fbp^2 -2 \ddot{\fb}  +\sum_iH_i^2=\sgb e^{\fb}, 
\label{6}
\eea
where $H_i=d(\ln a_i) /dt$, $t$ is the cosmic time, and we have
introduced the convenient ``shifted" variables
\beq
{\fb} = \phi- \ln \sqrt{-g},~~~~
\rb=  \r \sqrt{-g}, ~~~~ \pb_i=  p_i \sqrt{-g}, ~~~~
\sgb=  \sg \sqrt{-g}, ~~~~
\sqrt{-g}= \prod_i a_i.
\eeq

In order to solve the above system of $D+1$ equations, for the $2D+1$
variables $\{a_i, \phi, \r, p_i, \sg\}$, we now need  $D$ ``equations
of state" relating $p_i$ and $\sg$ to the energy density of the sources. 
In a complete, and fully realistic scenario, including all the relevant
matter fields, $p_i$ and $\sg$ are in general complicated functions of
$\r$, with time-dependent coefficients. However, since we are mainly
interested in the graceful exit, here we shall restrict our discussion
to the transition regime, where the back-reaction of the quantum
fluctuations is expected to give the dominant contribution to $\Ga$, and
we shall assume a simple ``barotropic" equation of state, 
\beq
p_i= \ga_i \r, ~~~~~~~~~~~~~~ \sg = \ga_0 \r,
\label{8}
\eeq
where $\ga_i, \ga_0$ are $D$ constant parameters specific to the given
model of matter fields and of their quantum fluctuations. 

In that case the system of equations (\ref{6}) can be integrated
exactly, following the method developed in \cite{1} and already
applied to various classes of homogeneous  backgrounds \cite{21}.
By introducing a new (dimensionless) time-coordinate $x$, such that
\beq
{1\over 2}\rb ={1\over L} {dx\over dt}
\label{9}
\eeq
($L$ is a constant parameter, with dimension of length), the equations
(\ref{6}) can be integrated a first time to give:
\bea
&&
\fb ' =-2(1+\ga_0) {(x+x_0)\over D(x)},
~~~~~~~~~~~~~~~~(1+\ga_0)\not=0, \label{10}\\
&&
{a_i'\over a_i}=2(\ga_i+\ga_0) {(x+x_i)\over D(x)},
~~~~~~~~~~~~~~~~~(\ga_i+\ga_0)\not=0
\label{11}
\eea
(a prime denotes differentiation with respect to $x$). Here  $x_i$ and
$x_0$ are $D$ integration constants, and $D(x)$ is a quadratic form
related to $\rb$ by
\beq
L^2 \rb e^{-\fb} = D(x) \equiv (1+\ga_0)^2 (x+x_0)^2 -
\sum_i (\ga_i+\ga_0)^2 (x+x_i)^2.
\label{12}
\eeq

The above equations hold for $(1+\ga_0) \not=0$, and
$(\ga_i+\ga_0)\not=0$. If $(1+\ga_0) =0$, however, eq. (\ref{10}) is  to
be replaced by
\beq
\fb ' =-2{x_0\over D(x)},
~~~~~~~~~~~~~~~~~(1+\ga_0)=0, 
\label{13}
\eeq
and the quadratic form becomes 
\beq
D(x)= x_0^2 -
\sum_i (\ga_i+\ga_0)^2 (x+x_i)^2.
\label{14}
\eeq
If  instead $(1+\ga_0)\not =0$, but $(\ga_i+\ga_0)=0$ for $i=1,2,\dots
n$, then  the first $n$ equations in (\ref{11}) are to be replaced by
\beq
{a_i' \over a_i}  =2{x_i\over D(x)},
~~~~~~~~~~~(\ga_i+\ga_0)=0, ~~~~~~~~~~~i= 1,2 \dots n,
\label{15}
\eeq
and the quadratic form becomes 
\beq
D(x)= (1+\ga_0)^2 (x+x_0)^2 - \sum_{i=1}^n x_i^2 -
\sum_{i=n+1}^D (\ga_i+\ga_0)^2 (x+x_i)^2.
\label{16}
\eeq
In both cases, $L^2 \rb e^{-\fb} = D(x) $. 

To discuss the possibility of graceful exit,  we should now separately
consider the various possibilities for the values of $(1+\ga_0)$
and $(\ga_i+\ga_0)$. However, as shown by a detailed analysis, in the
case   $(1+\ga_0)=0$  the curvature cannot be regular everywhere: 
even if $D(x)$ is always non-zero, the curvature necessarily blows up at
$x \ra \pm \infty$. On the other hand, if $(\ga_i+\ga_0)=0$,  the
condition of smooth curvature turns out to be incompatible with the
condition of smooth energy density, $|\r|<\infty$. We shall thus
concentrate, in the following discussion,  on the set of equations
(\ref{10}--\ref{12}), and we shall introduce the convenient definitions:
\bea
&&
D(x)= \a x^2+bx +c, ~~~~~~~~~~~~~~~~~~~~~~~~~~
\a= (1+\ga_0)^2 - \sum_{i}(\ga_i+\ga_0)^2, 
\nonumber\\ &&
b= 2(1+\ga_0)^2 x_0- 2\sum_{i}(\ga_i+\ga_0)^2x_i, ~~~~~~~
c= x_0^2(1+\ga_0)^2 - \sum_{i}(\ga_i+\ga_0)^2x_i^2. 
\label{17}
\eea

A necessary condition for for the existence of smooth solutions is the
absence of zeros in the quadratic form $D(x)$. When the background is
isotropic, i.e. $\ga_i$ and $x_i$ have the same values for all the $D-1$
spatial directions, then the discriminant of $D(x)$ is always
non-negative,
\beq
\Da = b^2-4 \a c= 4(D-1) (1+\ga_0)^2(\ga_i+\ga_0)^2(x_i-x_0)^2
\geq 0,
\label{18}
\eeq
and $D(x)$ necessarily has zeros on the real axis, correponding to
singularities both in the curvature and in the dilaton kinetic energy. 
A negative value of $\Da$ can be obtained, however, when 
$\ga_i$ and $x_i$ have different values in different directions. 
Here is why anisotropy is needed, for a graceful exit. 

To illustrate this possibility we shall consider a simple example of
background, in which the spatial geometry is factorizable as the direct
product of two conformally flat manifolds with $d$ and $n$ dimensions,
respectively, so that we can set:
\bea
&&
a_i=a_1, ~~~~~~~~~ \ga_i=\ga_1, ~~~~~~~~~
x_i=x_1, ~~~~~~~~~ i=1, \dots d, 
\nonumber\\
&&
a_i=a_2, ~~~~~~~~~ \ga_i=\ga_2, ~~~~~~~~~
x_i=x_2, ~~~~~~~~~ i=d+1, \dots d+n. 
\label{19}
\eea
Also, we shall choose a convenient set of integration constants, such
that the linear term in the quadratic form (\ref{17}) disappears. For
instance:
\beq
x_0=0, ~~~~~~~~~~~~
x_1= -x_2{n (\ga_2+\ga_0)^2\over d (\ga_1+\ga_0)^2}.
\label{20}
\eeq
It turns out that $c<0$, and that the absence of zeros in $D(x)$ can be
avoided, $\Da=-4 \a c <0$, provided 
\beq
\a = (1+\ga_0)^2-d(\ga_1+\ga_0)^2 -n (\ga_2+\ga_0)^2 <0.
\label{21}
\eeq

If this condition is satisfied then $D(x) <0$ everywhere, and this
implies, through eq. (\ref{12}), $\r<0$ (note that the result $D(x)<0$ in
the absence of zeros is independent from the particular choice $b=0$). 
As discussed before, this agrees with our expectation that during the
exit the dominant  contribution to the gravitational sources should come
from the back-reaction of the quantum fluctuations outside the horizon,
when their effective energy density is indeed negative \cite{15,16}. 
We stress again that such a negative energy density goes to zero at
large times (well inside the post-big bang regime), when the horizon
becomes larger and larger and all modes of the quantum fluctuations
re-enter inside the horizon, giving rise to the well known phenomenon
of cosmological particle production. The energy density thus
asymptotically switches to a positive regime, dominated by the
contribution of the effective stress tensor of the produced radiation
\cite{20}. Such an asymptotic regime will not be considered in this
paper, as here we are mainly interested in the discussion of the exit,
and we shall concentrate our attention on the transition regime where
the backreaction of particle production is negligible. 

When the conditions (\ref{19}--\ref{21}) are satisfied, the integration
of eqs. (\ref{10},\ref{11}) leads to the exact solution
\bea
&&
e^{\fb}= e^{\phi_0}\left|D(x)\right|^{-{1+\ga_0\over \a}},
~~~~~~~~~~~~ \rb= - {e^{\phi_0}\over L^2}
\left|D(x)\right|^{1-{1+\ga_0\over \a}},  \nonumber\\ &&
{a_i} = a_{i0} E_i (x)\left|D(x)\right|^{{\ga_i+\ga_0\over
\a}},  ~~~~~
E_i(x)= \exp\left[{2x_i (\ga_i+\ga_0)\over \sqrt{\a c}} \tan^{-1} \left(\a x
\over \sqrt{\a c}\right) \right], ~~~~~ i=1,2,
\label{22}
\eea
where $\phi_0$ and $a_{i0}$ are integration constants. Using eq.
(\ref{9}) we can then obtain the corresponding Hubble parameters $H_i=
(a_i'/a_i)(dx/dt)$, and the dilaton kinetic energy $\dot \phi= \fbp
+dH_1+nH_2$: 
\bea 
&&
H_1=  {e^{\phi_0}\over
L}(\ga_1+\ga_0){(x+x_1)}\left|D(x)\right|^{-{1+\ga_0\over \a}}, ~~~~~
H_2=  {e^{\phi_0}\over
L}(\ga_2+\ga_0){(x+x_2)}\left|D(x)\right|^{-{1+\ga_0\over \a}},
\nonumber\\ 
&& 
\dot \phi ={e^{\phi_0}\over L}\left|D(x)\right|^{-{1+\ga_0\over
\a}}\left[- (1+\ga_0)x+d (\ga_1+\ga_0)(x+x_1)+n
(\ga_2+\ga_0)(x+x_2)\right] 
\label{23} 
\eea
(for $\r<0$, it is convenient to choose $L<0$, so that $dx/dt >0$).
Finally, by rescaling $\fb, \rb$ through the explicit solutions for the
scale factors, we can also obtain the evolution of the non-shifted
variables:
\bea
&&
e^{\phi}={e^{\phi_0}}a_{10}^d a_{20}^n
E_1^d(x) E_2^n(x) \left|D(x)\right|^{-[(1+\ga_0)-d(\ga_1+\ga_0)-n 
(\ga_2+\ga_0)]/\a}, \nonumber\\
&&
\r=- {e^{\phi_0}\over L^2}a_{10}^{-d} a_{20}^{-n}
E_1^{-d}(x) E_2^{-n}(x)
\left|D(x)\right|^{1-[(1+\ga_0)+d(\ga_1+\ga_0)+n  (\ga_2+\ga_0)]/\a}. 
\label{24}
\eea

The above exact solution satisfies the condition (\ref{21}), which is
necessary for a model for graceful exit, but non sufficient. In addition,
we have to impose that the curvature and the dilaton kinetic energy of
eq. (\ref{23}), toghether with the effective string coupling $e^\phi$, 
are bounded everywhere. This requires, respectively:
\beq
2(1+\ga_0)<\a, ~~~~~~~~~
(1+\ga_0)-d(\ga_1+\ga_0)-n (\ga_2+\ga_0)<0.
\label{25}
\eeq 
The energy density $\r$ of eq. (\ref{24}) also should be bounded and, in
particular, should go asymptotically to zero at large times, to be
consistently interpreted as the contribution of the quantum
back-reaction. This imposes the condition
\beq
(1+\ga_0)+d(\ga_1+\ga_0)+n (\ga_2+\ga_0)<\a. 
\label{27}
\eeq
Finally, for possible applications to a realistic scenario, our
anisotropic background should contain, in its final configuration, $d$
expanding and $n$ contracting dimensions. This requires (see the
solutions for $a_i$ in  eq. (\ref{22})): 
\beq
\ga_1+\ga_0<0, ~~~~~~~~~~~~~ \ga_2+\ga_0>0.
\label{28}
\eeq
A consistent and successful  model of graceful exit should
satisfy the whole set of conditions (\ref{21}), (\ref{25}--\ref{28}). 

A detailed analysis of the above inequalities shows that there 
is a region of non-zero extension in the space of the parameters $\ga_i,
\ga_0$ for which all the conditions are satisfied. This means that, if the
back-reaction  generated by the quantum fluctuations is appropriate, a
model of graceful exit can be implemented even in the context of the
low-energy string effective action, without higher-derivative
corrections. 

In order to check  our analytical results, we have numerically
integrated the string cosmology equations (\ref{6}), using directly the
cosmic time variable.  Such equations, when applied to  the factorized
configuration (\ref{19}), are equivalent to a
system of  four independent equations for the four variables $H_i,
\phi,\r$ ($i=1,2$): 
\bea
&&
\dot H_i -H_i \left(\dot\phi -dH_1-nH_2\right)={1\over 2} \r 
(\ga_i+\ga_0)e^\phi, \nonumber\\
&&
\left(\dot\phi -dH_1-nH_2\right)^2-2\left(\ddot \phi -d \dot H_1 -n
\dot H_2\right) +dH_1^2+nH_2^2 =\ga_0\r e^\phi,
\nonumber\\
&&
\dot \r +dH_1 (1+\ga_1)\r+nH_2(1+\ga_2)\r+ \ga_0 \r \dot \phi=0.
\label{29}
\eea
We have used, for the numerical integration, the following  set of
parameters: 
\beq
d=3,~~~~ n=6, ~~~~\ga_0=-3.25, ~~~~\ga_1=2.25,
 ~~~~\ga_2=3.85,
\label{30}
\eeq
satisfying all the inequalities  (\ref{21}), (\ref{25}--\ref{28}).  We
have imposed, as initial conditions, a small and negative energy
density, $\rho_{in}<0$, and a small but increasing dilaton,  $\dot
\phi_{in}>0$. We have also restricted the initial conditions to lie on the
 trajectory of our analytical solution (\ref{23}), (\ref{24}), using the
fact that, at fixed $x=0$, the choice of parameters (\ref{30}) leads to
the relations:
\beq
H_1(0)= 1.2 H_2(0), ~~~~~~~~~~~~~~~~
\dot \phi (0)= 8 H_1(0).
\label{30a}
\eeq
The full set of initial conditions is further restricted by the 
 Hamiltonian constraint (first of eqs. (\ref{6})) as 
\beq
\left(\dot\phi -dH_1-nH_2\right)^2 -dH_1^2-nH_2^2= \r e^\phi.
\label{31}
\eeq
The results of the numerical integration are shown in Fig. 1. 

\begin{figure}[t]
\begin{center}
\includegraphics{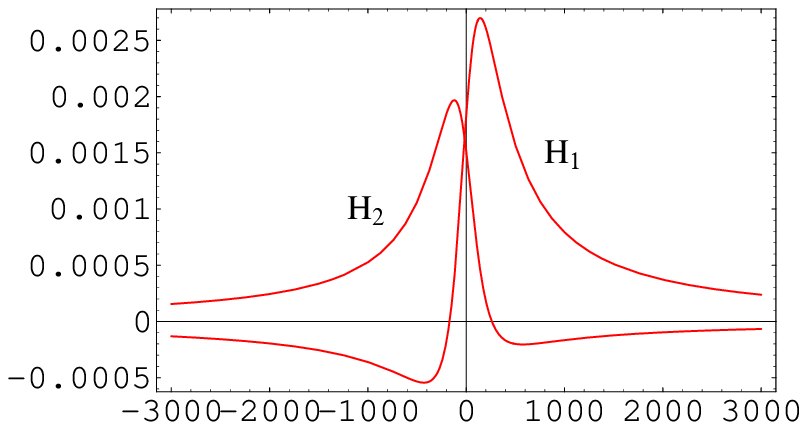}
\includegraphics{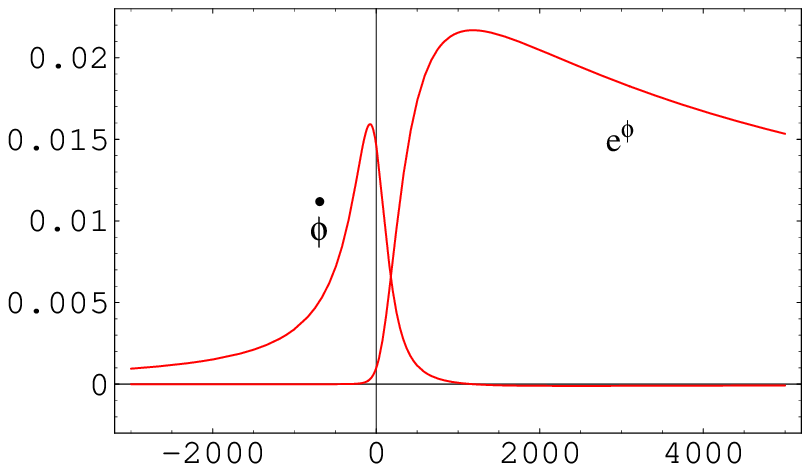}
\includegraphics{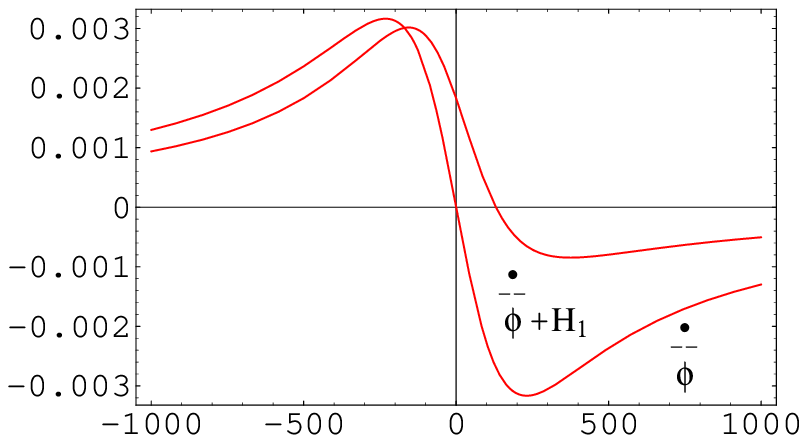}
\includegraphics{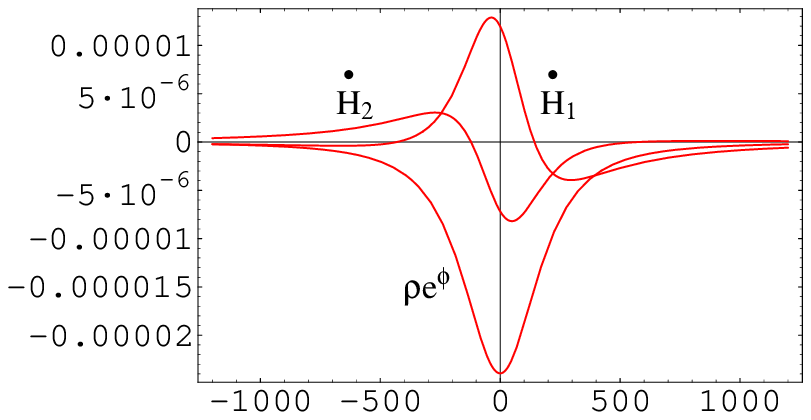}
\vskip 5mm
\caption{\sl 
The plots show the evolution  in cosmic time of $H_i$, $\dot H_i$, $\dot
\phi$,  $\fbp$, $e^\phi$,    $\r e^\phi $, obtained through a numerical
integration of eqs.  (\ref{29}), with the set of parameters given in eq.
(\ref{30}), and with the following initial conditions (satisfying eqs.
(\ref{30a}), (\ref{31})), imposed at $t=0$: $H_1=0.00182772$, 
$H_2=0.0015231$, $\dot \phi=0.0146218$,  $\phi=-6.91139$,  
$\rho=-0.24028$.}  
\end{center} \end{figure}

In the example illustrated in Fig. 1 the background undergoes a smooth 
and homogeneous evolution from a pre-big bang phase in which the
curvature and the dilaton are increasing, to a post-big bang phase in
which the curvature and the dilaton are decreasing ($\dot \phi \ra 0$
from negative values as $t \ra +\infty$). The final post-big bang
configuration is characterized by $H_1>0$, $H_2<0$ for $t \ra+\infty$,
and thus describes $3$ expanding and $6$ contracting  spatial
dimensions, as appropriate to a phase of dynamical dimensional
reduction in a superstring theory context ($D=1+d+n=10$). Also, the final
configuration satisfies all the prescribed conditions \cite{10} for a
successful exit, i.e. $\fbp <0$, $\fbp <-H_1$  as $t \ra+\infty$. 
The negative energy density of the sources (not shown in the picture) is
bounded and goes to zero, far from the transition regime, as appropriate
to the back-reaction generated by the quantum fluctuations outside the
horizon. Finally, all the curvature terms ($H_i^2, \dot \phi^2, \dot H_i$)
appearing in the equations, including the source term $e^\phi \r$,
remains much smaller than one in string units, as appropriate to an
action describing low-energy dynamics. 

\begin{figure}[t]
\begin{center}
\includegraphics{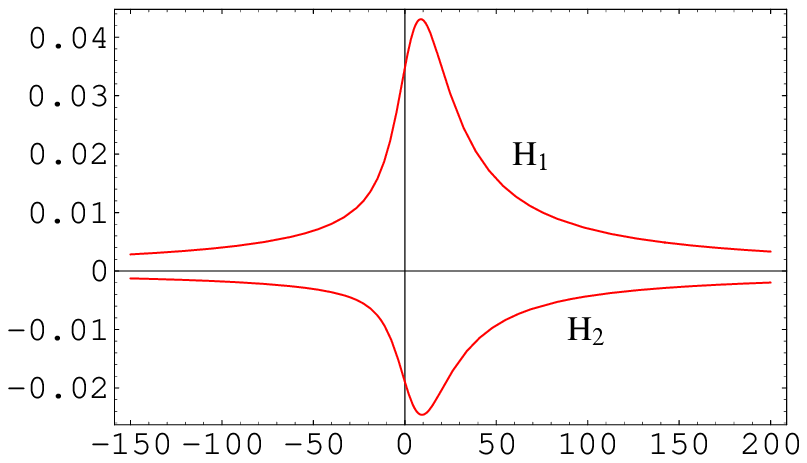}
\includegraphics{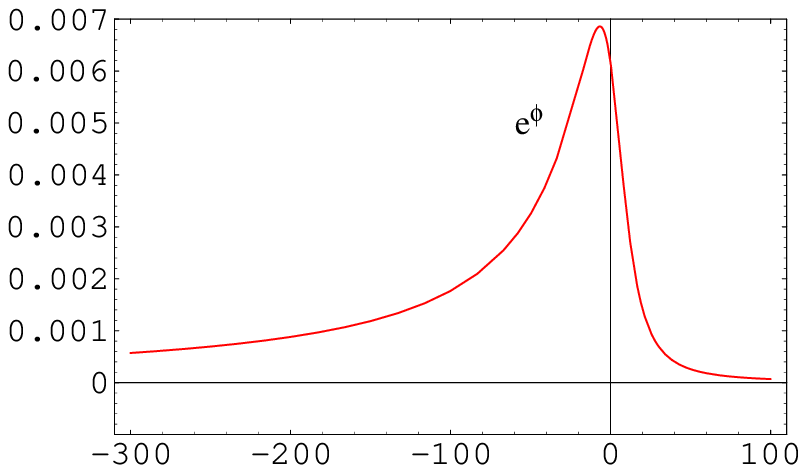}
\includegraphics{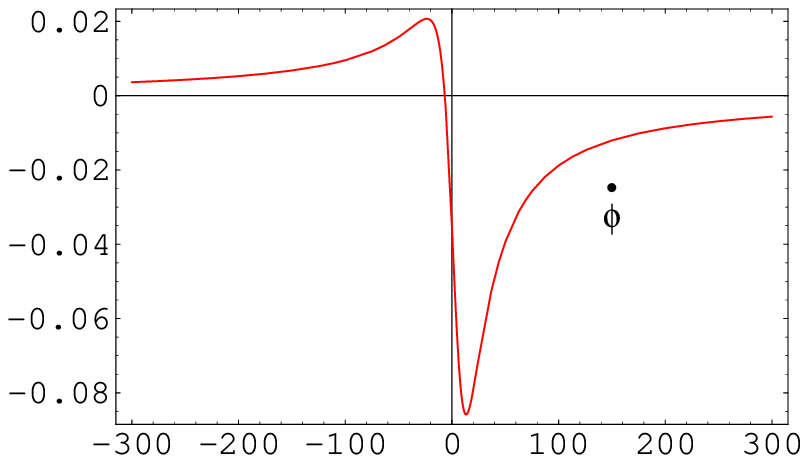}
\includegraphics{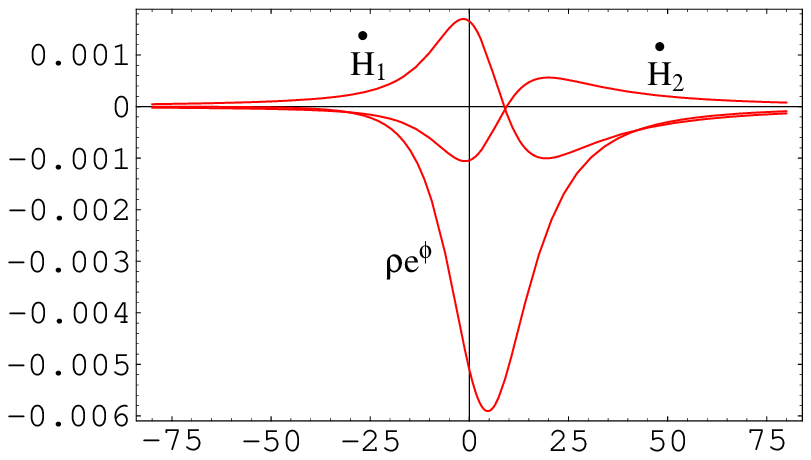}
\vskip 5mm
\caption{\sl 
The plots show the evolution  in cosmic time of $H_i$, $\dot H_i$, $\dot
\phi$, $e^\phi$,    $\r e^\phi $, obtained through a numerical
integration of eqs.  (\ref{29}), with the set of parameters given in eq.
(\ref{30}), and with the following initial conditions (satisfying eq.
 (\ref{31})) imposed at $t=-10$: $H_1=0.02$, $H_2=-0.01$, 
$\dot \phi=0.01$, $\phi=-5$, $\rho=-0.25230237$.}  
\end{center}
\end{figure}

It is important to stress that the exact analytical solution (\ref{23}),
(\ref{24}), reproduced numerically in Fig. 1, is only a special example of 
smooth transition corresponding to the particular choice of integration
constants given in eq. (\ref{20}). In general, other smooth
configurations are allowed, including also the case of a monotonic
evolution of the ``external" and ``internal" scale factors $a_1$ and
$a_2$. This possibility is illustrated in Fig. 2, in which we report the
results of a numerical integration of eqs. (\ref{29}), with the same set 
of parameters given in eq. (\ref{30}), and with initial conditions
satisfying the Hamiltonian constraint (\ref{31}) but {\em not} the
constraints (\ref{30a}), typical of our particular analytical example. 
The numerical example of Fig. 2, in particular, describes a smooth
transition in which the three external dimensions evolve from
accelerated to decelerated expansion, while the six internal
dimensions from accelerated to decelerated contraction. 
The simultaneous flip in sign of $\dot H_1$, $\dot H_2$, illustrated in the
picture, marks the end of the phase of pre-big bang inflation and the
beginning of the standard decelerated regime.

In conclusion,  the combined effect of anisotropy (physically associated
to the dimensional reduction) and of a negative energy density
(physically associated to the quantum back-reaction) seem to be able
to trigger an efficient and graceful exit from the pre-big bang regime,
even at small curvatures, at least for an appropriate range of
parameters characterizing the source stress tensor. The toy model
that we have presented in this paper, to illustrate the joint effects of
anisotropy and back-reaction, is not intended, of course, to
represent an exhaustive and fully realistic picture of the complete
transition to the post-big bang regime -- other effects, like $\a'$
corrections, can in principle become important near the transition
regime. In addition, at late times,  a dilaton potential is expected to
be added, and to play a possible significant role for the
dilaton evolution. Also, at late times, the (positive) radiation energy
density, due to particle production effects, is expected to isotropize the
background and possibly contribute to dilaton stabilization, as discussed
in \cite{20}.   The conditions (\ref{21}), (\ref{25}--\ref{28})
determined in this paper, however, can be applied to various models of
(classical or quantum) sources, in the transition regime, to obtain ``a
priori" indications on the effective back-reaction of their fluctuations
outside the horizon, and on their possible ability of driving a smooth
evolution from the string perturbative vacuum to our present
cosmological configuration.

\acknowledgments
It is a pleasure to thank Enrico Onofri for useful suggestions
concerning the numerical integrations.
We are also grateful to Gabriele Veneziano for many helpful comments
and discussions.

\end{document}